\documentclass[12pt]{article}
       \usepackage{amsmath,amsthm,upref}
       \usepackage{amssymb,amsbsy}
       \usepackage{amsfonts}
\usepackage{cite}
\usepackage[english,russian]{babel}
\usepackage[cp1251]{inputenc}
\usepackage[left=2.3cm,body={17.1cm,23.5cm},top=2.5cm]{geometry}
\author{{\bf С.А. Дуплий}\\
{\it Физико-технический факультет,}\\ 
{\it Харьковский национальный университет им. В.Н. Каразина,}\\
{\it пл. Свободы, 4, Харьков, 61022, Украина}\\
 E-mails: duplij@math.rutgers.edu,\; sduplij@gmail.com\\
 URL: http://math.rutgers.edu/\~{}duplij
}

\title{{\bf Частичный гамильтонов формализм, многовременная динамика и сингулярные теории} }

\date{20 апреля 2013 г.}

\begin{document}
\thispagestyle{empty}
\begin{center}
\ \textbf{Partial Hamiltonian formalism, multi-time dynamics and
singular theories} 

{\small \textbf{Steven Duplij} }

{\small \textit{Department of Physics and Technology}, \textit{V.N. Karazin
Kharkov National University}\\[0pt]
\textit{Svoboda Sq. 4, Kharkov 61022, Ukraine }\\[0pt]
\textrm{E-mails: duplij@math.rutgers.edu, sduplij@gmail.com\; URL: http://math.rutgers.edu/\~{}%
duplij}}
\end{center}
{\centerline {\bf Abstract}}

{\small \noindent We formulate singular (with degenerate Lagrangians) classical theories (for
clarity, in local coordinates) \textit{without involving constraints}. First,
we recall the standard action principle (for pedagogical reasons and in order
to establish notation). Then, applying it to the action (27), we develop a
\textit{partial }(in the sense that \textit{not all} velocities are transformed to
momenta) Hamiltonian formalism in an initially reduced phase space (with
canonical coordinates $q_{i},p_{i}$, where the number $n_{p} \leq n$ of momenta
$p_{i}$, $i=1,\ldots,n_{p}$, see (17), is \textit{arbitrary}, where $n$ is the dimension of the configuration space) in terms of a
partial Hamiltonian $H_{0}(q_{i},p_{i},q^{\alpha},\dot{q}^{\alpha})$, see (18), and
$(n-n_{p})$ additional Hamiltonians $H_{\alpha}(q_{i},p_{i},q^{\alpha},\dot
{q}^{\alpha})$, $\alpha=n_{p}+1,\ldots,n$, see (20) (\textit{instead} of the
remaining momenta $p_{\alpha}$ defined in the standard full Hamiltonian
formalism (6)). In this way we obtain $(n-n_{p}+1)$ Hamilton-Jacobi equations
(25)-(26) which fully determine the dynamics. The equations of motion are first-order differential equations (33)-(34) with respect to the canonical
coordinates $q_{i},p_{i}$ and second-order differential equations (35) in the
noncanonical coordinates $q_{\alpha}$ (which \textit{have no} corresponding
momenta). In the partial Hamiltonian formalism (which describes the same
dynamics as the Lagrange equations of motion (3)), the number of momenta
$n_{p}\leq n$ is arbitrary. The limit cases $n_{p}=n$ and $n_{p}=0$ correspond
to the standard Hamiltonian and Lagrangian dynamics (discussed in (37)-(41)), respectively.

If the Hamiltonians $H_{0}(q_{i},p_{i},q^{\alpha})$, $H_{\alpha}(q_{i}%
,p_{i},q^{\alpha})$ do not depend of the noncanonical velocities $\dot
{q}_{\alpha}$ (conditions (42)), then the second-order differential equations
(35) become \textit{purely algebraic} equations (43) with respect to $\dot
{q}_{\alpha}$. In this case we can interpret the noncanonical coordinates
$q_{\alpha}$ as additional times by (45), such that the partial Hamiltonian
formalism becomes equivalent to multi-time dynamics with action (51) and
equations of motion (53)-(54) with additional (integrability) conditions
for the Hamiltonians (55).

The independence of the Hamiltonians $H_{0}(q_{i},p_{i},q^{\alpha})$,
$H_{\alpha}(q_{i},p_{i},q^{\alpha})$ of the noncanonical velocities $\dot
{q}_{\alpha}$ (conditions (42)) is satisfied in singular theories (with
degenerate Lagrangian, in the sense that the determinant of the Hessian matrix is zero), in
which case the rank $r_{W}$ of the Hessian is less than or equal to the number of
momenta: $r_{W}\leq n_{p}$, see (58). If we choose $n_{p}>r_{W}$, then we obtain
$(n_{p}-r_{W})$ primary (and perhaps, higher-level) constraints (as in the Dirac
theory), but if we assume the equality $n_{p}=r_{W}$, then there will be \textit{no
constraints at all}. The $(n-r_{W})$ equations for \textit{the same number} of
$(n-r_{W})$ noncanonical velocities $\dot{q}_{\alpha}$ (62) constitute a
standard system of linear algebraic equations, but \textit{not constraints},
because we do not define the \textquotedblleft extra\textquotedblright%
\ momenta $p_{\alpha}$, and the dynamics is fully described \textit{without
them} by the equations of motion (60)-(62).

Classification of singular theories can be made by the analysis of the linear
algebraic system (62) in terms of the rank of the tensor $F_{\alpha\beta}$
(63). If its rank is full, i.e., equal to $(n-r_{W})$, then we can solve the
system (62), and there will be no arbitrary parameters (gauge degrees of
freedom) in the theory; in other cases, some of the noncanonical velocities
$\dot{q}_{\alpha}$ will remain arbitrary, which is a sign of gauge theory.
In both cases we define new antisymmetric brackets (69) and (80) (which
govern time evolution of physical variables (70) and (83)) and present the
equations of motion in the Hamilton-like form, (67)-(68) and (81)-(82), respectively.

Finally, we clarify the origin of the Dirac constraints in our framework: if
we define $(n-r_{W})$ \textquotedblleft extra\textquotedblright\ dynamical
variables, that is to say, the momenta $p_{\alpha}$ by (86), then we obtain the
standard primary constraints (87) and the equations of motion (90)-(91) in
terms of the Dirac total Hamiltonian (88). In this case our new brackets (69)
and (80) will transform into the Dirac bracket.

At the end of the paper, quantization is discussed briefly.

}

\maketitle

\bigskip

\noindent
{\small \noindent В работе п%
риводится ф%
ормулировка
сингулярных
теорий (с выр%
ожденными л%
агранжианам%
и) без привле%
чения связе%
й. Строится ч%
астичный га%
мильтонов ф%
ормализм в р%
едуцированн%
ом фазовом п%
ространстве
(с произволь%
ным количес%
твом импуль%
сов). Уравнен%
ия движения 
становятся 
дифференциальными
уравнениями
первого пор%
ядка и совпа%
дают с уравн%
ениями мног%
овременной 
динамики пр%
и определен%
ых условиях, 
которыми в с%
ингулярных 
теориях явл%
яется совпа%
дение числа 
обобщенных 
импульсов с 
рангом матр%
ицы гессиан%
а. Неканонич%
еские обобщ%
енные скоро%
сти удовлет%
воряют сист%
еме алгебра%
ических лин%
ейных уравн%
ений, что зад%
ает соответ%
ствующую кл%
ассификацию
сингулярных
теорий (кали%
бровочные и 
некалиброво%
чные). Для опи%
сания эволю%
ции во време%
ни физическ%
их величин в%
водится нов%
ая антисимм%
етричная ск%
обка (аналог 
скобки Пуас%
сона). Показа%
но, каким обр%
азом при рас%
ширении фаз%
ового прост%
ранства поя%
вляются свя%
зи, при этом н%
овая скобка 
переходит в 
скобку Дира%
ка. Кратко об%
суждается к%
вантование.}

\bigskip

 \par \noindent
 {\small {\bf Ключевые
слова}:\hskip 3pt
действие, вырожденный лагранжиан, гамильтониан, гессиан, 
ранг, уравнение Гамильтона-Якоби, многовременная динамика, скобка
 } 

\bigskip 

\newpage

\section{Введение}

Многие
современные
физические
модели
являются
калибровочными
теориями (см.,
например, \cite{weinberg123}),
которые на
классическом
уровне
описываются
сингулярными
(вырожденными)
лагранжианами
\cite{car6}. Обычно, для
последовательного
квантования
используется
гамильтонов
формализм,
переход к
которому для
сингулярных
теорий
нетривиален
(из-за того,
что
невозможно
непосредственно
применить
преобразование
Лежандра \cite{tul,tul/urb})
и требует
дополнительных
построений
\cite{men/tul,mar/men/tul}.
Основная
трудность
заключается
в появлении
дополнительных
соотношений
между
динамическими
переменными,
которые
называются
связями \cite{dirac}.
Далее,
важным
является
разрешение
этих связей
и выделение
физического
подпространства
(редуцированного
фазового
пространства),
на котором
можно
последовательно
проводить
процедуру
квантования
\cite{git/tyu,hen/tei}.
Несмотря на
широкое
применение
теории
связей \cite{sundermeyer,reg/tei},
она сама не
лишена
внутренних
противоречий
и проблем \cite{pon3,mik/zan}.
Поэтому
имеет смысл
пересмотреть
сам
гамильтонов
формализм
для
сингулярных
теорий с
вырожденными
лагранжианами
\cite{dup2009,dup2011}.

Целью данной
работы
является
описание
сингулярных
теорий без
помощи
связей.
Вначале
приводится
построение
частичного
гамильтонового
формализма,
в котором
для любой
лагранжевой
системы
строится
гамильтонова
система с
произвольным
и заранее
нефиксированным
количеством
импульсов
(произвольно
редуцированное
фазовое
пространство).
Соответствующая
система
уравнений
движения,
полученная
из принципа
наименьшего
действия,
содержит
производные
первого
порядка
канонических
переменных и
производные
второго
порядка
неканонических
обобщенных
координат.
При
определеных
условиях
уравнения
для
последних
становятся
дифференциально-алгебраическими
уравнениями
первого
порядка, и
такая
физическая
система
эквивалентна
многовременной
динамике.
Эти условия
реализуются
в теории с
вырожденными
лагранжианами,
если выбрать
число
импульсов,
которое
совпадает с
рангом
гессиана.
Тогда
уравнения
для
неканонических
обобщенных
скоростей
становятся
алгебраическими,
и динамика
определяется
в терминах
новых
скобок,
которые, как
и скобки
Пуассона,
являются
антисимметричными
и
удовлетворяют
тождеству
Якоби. В
приведенном
формализме
не возникает
никаких
дополнительных
соотношений
между
динамическими
переменными
(связей).
Показано,
что, если
расширить
фазовое
пространство
так, чтобы
импульсы
были
определены и
для
неканонических
обобщенных
скоростей,
то в
результате
этого и
появляются
связи, а
соответствующе
формулы
воспроизводят
теорию
связей
Дирака \cite{dirac}. 

Для
ясности
изложения мы
пользуемся
локальными
координатами
и
рассматриваем
системы с
конечным
числом
степеней свободы.

\section{Действие и
полный
гамильтонов
формализм}

Рассмотрим
динамическую
систему,
которая
может быть
определена в
терминах
обобщенных
координат $q^{A}\left(
t\right)  $, $A=1,\ldots,n$ (как
функций
времени) в
конфигурационном
пространстве
$Q_{n}$
размерности
$n$. Эволюция
динамической
системы, то
есть
траектория в
конфигурационном
пространстве
$Q_{n}$,
определяется
уравнениями
движения,
которые
представляют
собой
дифференциальные
уравнения
для
обобщенных
координат $q^{A}\left(
t\right)  $ и их
производных
по времени $\dot{q}%
^{A}\left(  t\right)  $, где $\dot{q}^{A}\left(  t\right)
\equiv dq^{A}\left(  t\right)  \diagup dt$,
которые
определяют
касательное
расслоение $TQ$
ранга $n$ (так
что
размерность
тотального
пространства
равна $2n)$ \cite{arnold}.
Здесь мы не
рассматриваем
системы с
высшими
производными
(см., например,
\cite{nak/ham,and/gon/mac/mas}).
Уравнения
движения
можно
получить с
помощью
различных
принципов
действия,
которые
отождествляют
реальную
траекторию с
требованием
экстремальности
некоторого
функционала
\cite{lanczos}.

В
стандартном
принципе
наименьшего
действия \cite{lanczos}
рассматривается
функционал%
\begin{equation}
S=\int_{t_{0}}^{t}L\left(  t^{\prime},q^{A},\dot{q}^{A}\right)  dt^{\prime
},\label{sl}%
\end{equation}
где
дифференцируемая
функция $L=L\left(  t,q^{A},\dot
{q}^{A}\right)  $ есть
лагранжиан,
а функционал
(на
экстремалях)
$S=S\left(  t,q^{A}\right)  $ ---
действие
динамической
системы как
функция
верхнего
предела $t$ (при
фиксированном
нижнем
пределе $t_{0}$).
Рассмотрим
бесконечно
малую
вариацию
функционала
(\ref{sl}) $\delta S=S\left(  t+\delta t,q^{A}+\delta q^{A}\right)  -S\left(
t,q^{A}\right)  $. Без
потери
общности
можно
считать, что
на нижнем
пределе
вариация
обращается в
нуль $\delta q^{A}\left(  t_{0}\right)  =0$, а
на верхнем
изменение
траектории
обозначим $\delta q^{A}%
$. Для
вариации $\delta S$
после
интегрирования
по частям
получаем\footnote{По
повторяющимся
нижним и
верхним
индексам
подразумевается
суммирование.
Индексы
внутри
аргументов
функций не
суммируются
и выписаны в
явном виде
для отличия
между собой
разных типов
переменных.}%
\begin{equation}
\delta S=\int_{t_{0}}^{t}\left(  \dfrac{\partial L}{\partial q^{A}}-\dfrac
{d}{dt^{\prime}}\left(  \dfrac{\partial L}{\partial\dot{q}^{A}}\right)
\right)  \delta q^{A}\left(  t^{\prime}\right)  dt^{\prime}+\dfrac{\partial
L}{\partial\dot{q}^{A}}\delta q^{A}+\left(  L-\dfrac{\partial L}{\partial
\dot{q}^{A}}\dot{q}^{A}\right)  \delta t,\label{ds}%
\end{equation}

Тогда из
принципа
наименьшего
действия $\delta S=0$
стандартным
образом
получаем
уравнения
движения
Эйлера-Лагранжа
\cite{lan/lif1}%
\begin{equation}
\dfrac{\partial L}{\partial q^{A}}-\dfrac{d}{dt}\left(  \dfrac{\partial
L}{\partial\dot{q}^{A}}\right)  =0,\ \ \ A=1,\ldots,n,\label{le}%
\end{equation}
которые
определяют
экстремали
при условии
закрепленных
концов $\delta q^{A}=0$ и $\delta
t=0$. Второе и
третье
слагаемые в
(\ref{ds})
определяют
полный
дифференциал
действия (на
экстремалях)
как функцию
$\left(  n+1\right)  $
переменных:
координат и
верхнего
предела
интегрирования
в (\ref{sl})%
\begin{equation}
dS=\dfrac{\partial L}{\partial\dot{q}^{A}}dq^{A}+\left(  L-\dfrac{\partial
L}{\partial\dot{q}^{A}}\dot{q}^{A}\right)  dt.\label{ds1}%
\end{equation}

Таким
образом, из
определения
действия (\ref{sl}) и
(\ref{ds1}) следует,
что%
\begin{equation}
\dfrac{dS}{dt}=L,\ \ \ \dfrac{\partial S}{\partial q^{A}}=\dfrac{\partial
L}{\partial\dot{q}^{A}},\ \ \ \dfrac{\partial S}{\partial t}=L-\dfrac{\partial
L}{\partial\dot{q}^{A}}\dot{q}^{A}.\label{sll}%
\end{equation}

В
гамильтоновом
формализме
каждой
координате $q^{A}
$ ставится в
соответствие
канонически
сопряженный
ей импульс $p_{A}$
по формуле%
\begin{equation}
p_{A}=\dfrac{\partial L}{\partial\dot{q}^{A}},\ \ \ A=1,\ldots,n.\label{p}%
\end{equation}
Если система
уравнений (\ref{p})
разрешима
относительно
всех
скоростей,
то можно
определить
гамильтониан
с помощью
преобразования
Лежандра%
\begin{equation}
H=p_{A}\dot{q}^{A}-L,\label{hp}%
\end{equation}
которое
определяет
отображение
между
касательным
и
кокасательным
расслоениями
$TQ_{2n}^{\ast}\rightarrow TQ_{2n}$ \cite{arnold}. В
правой части
(\ref{hp}) все
скорости
выражены
через
импульсы,
так что $H=H\left(  t,q^{A},p_{A}\right)
$ есть
функция в
фазовом
пространстве
(или на
кокасательном
расслоении
$TQ_{2n}^{\ast}$), то есть
зависящая от
$2n$
канонических
координат $\left(
q^{A},p_{A}\right)  $.
Поскольку в
стандартном
формализме
каждая
координата $q^{A}$
имеет свой
сопряженный
импульс $p_{A}$ по
формуле (\ref{p}),
назовем его
полным
гамильтоновым
формализмом
(или
гамильтоновым
формализмом
в полном
фазовом
пространстве
$TQ_{2n}^{\ast}$).

Тогда
дифференциал
действия (\ref{ds1})
может быть
также
записан в
полном
фазовом
пространстве%
\begin{equation}
dS=p_{A}dq^{A}-Hdt.\label{sp}%
\end{equation}
Поэтому для
частных
производных
действия
получаем%
\begin{equation}
\dfrac{\partial S}{\partial q^{A}}=p_{A},\ \ \ \dfrac{\partial S}{\partial
t}=-H\left(  t,q^{A},p_{A}\right)  ,\label{spd}%
\end{equation}
откуда
следует
дифференциальное
уравнение
Гамильтона-Якоби%
\begin{equation}
\dfrac{\partial S}{\partial t}+H\left(  t,q^{A},\dfrac{\partial S}{\partial
q^{A}}\right)  =0.
\end{equation}
Вариация
действия%
\begin{equation}
S=\int\left(  p_{A}dq^{A}-Hdt\right) \label{spi}%
\end{equation}
при
рассмотрении
координат и
импульсов
как
независимых
переменных и
интегрировании
по частям
приводят
стандартным
образом \cite{lan/lif1} к
уравнениям
Гамильтона в
дифференциальном
виде\footnote{Уравнения
(\ref{qp}) являются
условиями
замкнутости
дифференциальной
1-формы (\ref{sp})
(Пуанкаре-Картана)
\cite{arnold}.}%
\begin{equation}
dq^{A}=\dfrac{\partial H}{\partial p_{A}}dt,\ \ \ \ \ dp_{A}=-\dfrac{\partial
H}{\partial q^{A}}dt\label{qp}%
\end{equation}
для полного
гамильтонового
формализма
(то есть
динамическая
система
полностью
задана на $TQ_{2n}^{\ast}$).
Если ввести
(полную)
скобку
Пуассона%
\begin{equation}
\left\{  A,B\right\}  _{full}=\dfrac{\partial A}{\partial q^{A}}%
\dfrac{\partial B}{\partial p_{A}}-\dfrac{\partial B}{\partial q^{A}}%
\dfrac{\partial A}{\partial p_{A}},\label{abf}%
\end{equation}
то уравнения
(\ref{qp}) запишутся
в
стандартном
виде \cite{lan/lif1}%
\begin{equation}
dq^{A}=\left\{  q^{A},H\right\}  _{full}\ dt,\ \ \ \ dp_{A}=\left\{
p_{A},H\right\}  _{full}\ dt.\label{qpp}%
\end{equation}
Понятно, что
обе
формулировки
принципа
наименьшего
действия (\ref{sl}) и
(\ref{spi}) полностью
эквивалентны
(описывают
одну и ту же
динамику)
при
определениях
импульсов (\ref{p})
и
гамильтониана
(\ref{hp}).

\section{Частичный
гамильтонов
формализм}

Переход от
полного к
частичному
гамильтоновому
формализму и
многовременной
динамике
может быть
проведен с
помощью
следующей
аналогии \cite{lanczos}.
При изучении
параметрической
формы
канонических
уравнений и
действия (\ref{spi})
формально
вводилось
расширенное
фазовое
пространство
с
дополнительными
координатой
и импульсом%
\begin{align}
q^{n+1}  &  =t,\label{qn1}\\
p_{n+1}  &  =-H.\label{pn1}%
\end{align}
Тогда
действие (\ref{spi})
принимает
симметричный
вид и
содержит
только
первое
слагаемое \cite{lanczos}%
. Здесь мы
поступим
противоположным
образом и
зададимся
вопросом:
можно ли
наоборот,
уменьшить
количество
импульсов,
описывающих
динамическую
систему, то
есть
сформулировать
частичный
гамильтонов
формализм,
который был
бы
эквивалентен
(на
классическом
уровне)
лагранжевому
формализму?
Иными
словами,
можно ли
описать
систему с
начальным
действием (\ref{sl})
в
редуцированном
фазовом
пространстве,
построить в
нем
некоторый
аналог
действия (\ref{spi}), и
какие
дополнительные
условия для
этого
необходимы?

Оказывается,
что ответ на
все эти
вопросы
положительный
и приводит к
описанию
сингулярных
теорий (с
вырожденным
лагранжианом)
без введения
связей \cite{dup2009,dup2011}.

Определим
частичный
гамильтонов
формализм
так, что
сопряженный
импульс
ставится в
соответствие
не каждой $q^{A}$
по формуле (\ref{p}),
а только для
первых $n_{p}<n$
обобщенных
координат\footnote{Это
можно всегда
сделать
соответствующим
переобозначением
переменных.},
которые
назовем
каноническими
и обозначим
$q^{i}$, $i=1,\ldots n_{p}$.
Полученное
редуцированное
многообразие
$TQ_{2n_{p}}^{\ast}$
определяется
$2n_{p}$
редуцированными
каноническими
координатами
$\left(  q^{i},p_{i}\right)  $.
Остальные
обобщенные
координаты
(и скорости)
будем
называть
неканоническими
$q^{\alpha}$ (и $\dot{q}^{\alpha}$), $\alpha=n_{p}+1,\ldots n$,
они образуют
конфигурационное
подпространство
$Q_{n-n_{p}}$, которому
соответствует
касательное
расслоение
$TQ_{2\left(  n-n_{p}\right)  }$ (нижний
индекс
обозначает
соответствующую
размерность
тотального
пространства).
Таким
образам,
динамическая
система
теперь
задается на
прямом
произведении
(многообразий)
$TQ_{2n_{p}}^{\ast}\times TQ_{2\left(  n-n_{p}\right)  }$.

Для
редуцированных
обобщенных
импульсов
имеем%
\begin{equation}
p_{i}=\dfrac{\partial L}{\partial\dot{q}^{i}},\ \ \ i=1,\ldots,n_{p}%
.\label{pp}%
\end{equation}

Частичный
гамильтониан,
по аналогии
с (\ref{hp}),
определяется
частичным
преобразованием
Лежандра%
\begin{equation}
H_{0}=p_{i}\dot{q}^{i}+\dfrac{\partial L}{\partial\dot{q}^{\alpha}}\dot
{q}^{\alpha}-L,\label{hl0}%
\end{equation}
которое
определяет
(частичное)
отображение
$TQ_{2n_{p}}^{\ast}\times TQ_{2\left(  n-n_{p}\right)  }\rightarrow TQ_{2n}$
(см. (\ref{hp})). В (\ref{hl0})
канонические
обобщенные
скорости $\dot{q}^{i}$
выражены
через
редуцированные
канонические
импульсы $p_{i}$ с
помощью (\ref{pp}).
Для
дифференциала
действия (\ref{sp})
можно
записать%
\begin{equation}
dS=p_{i}dq^{i}+\dfrac{\partial L}{\partial\dot{q}^{\alpha}}dq^{\alpha}%
-H_{0}dt.\label{dss}%
\end{equation}

Введем
обозначение%
\begin{equation}
H_{\alpha}=-\dfrac{\partial L}{\partial\dot{q}^{\alpha}},\ \ \ \alpha
=n_{p}+1,\ldots n\label{ha}%
\end{equation}
и назовем
функции $H_{\alpha}$
дополнительными
гамильтонианами,
тогда%
\begin{equation}
dS=p_{i}dq^{i}-H_{\alpha}dq^{\alpha}-H_{0}dt.\label{spq}%
\end{equation}

Отметим, что
без второго
слагаемого
частичный
гамильтониан
(\ref{hl0})
представляет
собой
функцию
Раусса, в
терминах
которой
можно
переформулировать
уравнения
движения
Лагранжа \cite{lan/lif1}.
Однако
последовательная
формулировка
принципа
наименьшего
действия для
$S$ и
многовременной
динамики
сингулярных
систем \cite{dup2009}
естественна
в терминах
введенных
дополнительных
гамильтонианов
$H_{\alpha}$ (\ref{ha}).

Таким
образом, в
частичном
гамильтоновом
формализме
динамика
системы
полностью
определяется
не одним
гамильтонианом,
а набором из
$\left(  n-n_{p}+1\right)  $
гамильтонианов
$H_{0}$, $H_{\alpha}$, $\alpha=n_{p}+1,\ldots n$.

Действительно,
из (\ref{spq}) следует,
что частные
производные
действия $S=S\left(
t,q^{i},q^{\alpha}\right)  $ имеют
вид (см. (\ref{spd}))%
\begin{align}
\dfrac{\partial S}{\partial q^{i}}  & =p_{i},\\
\dfrac{\partial S}{\partial q^{\alpha}}  & =-H_{\alpha}\left(  t,q^{i}%
,p_{i},q^{\alpha},\dot{q}^{\alpha}\right)  ,\\
\dfrac{\partial S}{\partial t}  & =-H_{0}\left(  t,q^{i},p_{i},q^{\alpha}%
,\dot{q}^{\alpha}\right)  ,
\end{align}
откуда
получаем
систему $\left(  n-n_{p}+1\right)  $
уравнений
типа
Гамильтона-Якоби%
\begin{align}
\dfrac{\partial S}{\partial t}+H_{0}\left(  t,q^{i},\dfrac{\partial
S}{\partial q^{i}},q^{\alpha},\dot{q}^{\alpha}\right)   &  =0,\label{st}\\
\dfrac{\partial S}{\partial q^{\alpha}}+H_{\alpha}\left(  t,q^{i}%
,\dfrac{\partial S}{\partial q^{i}},q^{\alpha},\dot{q}^{\alpha}\right)   &
=0.\label{sq}%
\end{align}

Теперь, на
прямом
произведении
$TQ_{2n_{p}}^{\ast}\times TQ_{2\left(  n-n_{p}\right)  }$
действие
имеет вид%
\begin{equation}
S=\int\left(  p_{i}dq^{i}-H_{\alpha}dq^{\alpha}-H_{0}dt\right)  .\label{spq1}%
\end{equation}
Варьирование
в (\ref{spq1})
производится
независимо
по $2n_{p}$
редуцированным
каноническим
координатам
$q^{i},p_{i}$ и по $\left(  n-n_{p}\right)  $
неканоническим
обобщенным
координатам
$q^{\alpha}$. В
предположении,
что вариации
$\delta q^{i}$, $\delta p_{i}$, $\delta q^{\alpha}$ на
верхнем и
нижнем
пределах
зануляются,
после
интергрирования
по частям
для вариации
действия (\ref{spq1})
получаем%
\begin{align}
\delta S  &  =\int\delta p_{i}\left[  dq^{i}-\dfrac{\partial H_{0}}{\partial
p_{i}}dt-\dfrac{\partial H_{\beta}}{\partial p_{i}}dq^{\beta}\right]
+\int\delta q^{i}\left[  -dp_{i}-\dfrac{\partial H_{0}}{\partial q^{i}%
}dt-\dfrac{\partial H_{\beta}}{\partial q^{i}}dq^{\beta}\right]  +\nonumber\\
&  \int\delta q^{\alpha}\left[  \dfrac{\partial H_{\alpha}}{\partial\dot
{q}^{\beta}}d\dot{q}^{\beta}+\dfrac{\partial H_{\alpha}}{\partial q^{i}}%
dq^{i}+\dfrac{\partial H_{\alpha}}{\partial p_{i}}dp_{i}+\dfrac{d}{dt}\left(
\dfrac{\partial H_{0}}{\partial\dot{q}^{\alpha}}+\dfrac{\partial H_{\beta}%
}{\partial\dot{q}^{\alpha}}\dot{q}^{\beta}\right)  dt+\right. \nonumber\\
&  \left.  +\left(  \dfrac{\partial H_{\alpha}}{\partial q^{\beta}}%
-\dfrac{\partial H_{\beta}}{\partial q^{\alpha}}\right)  dq^{\beta}+\left(
\dfrac{\partial H_{\alpha}}{\partial t}-\dfrac{\partial H_{0}}{\partial
q^{\alpha}}\right)  dt\right]  .\label{dst}%
\end{align}
Уравнения
движения для
частичного
гамильтонового
формализма
можно
получить из
принципа
наименьшего
действия $\delta S=0$.
Принимая во
внимание тот
факт, что
вариации $\delta q^{i}$,
$\delta p_{i}$, $\delta q^{\alpha}$
независимы,
коэффициенты
при них
(каждая
квадратная
скобка в (\ref{dst}))
обращаются в нуль.

Введем
скобки
Пуассона для
двух функций
$A$ и $B$ на
редуцированном
фазовом
пространстве%
\begin{equation}
\left\{  A,B\right\}  =\dfrac{\partial A}{\partial q^{i}}\dfrac{\partial
B}{\partial p_{i}}-\dfrac{\partial B}{\partial q^{i}}\dfrac{\partial
A}{\partial p_{i}}.\label{ab}%
\end{equation}
Тогда,
подставляя
$dq^{i}$ и $dp_{i}$ из
первой
строки (\ref{dst}) во
вторую
строку,
получаем
уравнения
движения на
$TQ_{2n_{p}}^{\ast}\times TQ_{2\left(  n-n_{p}\right)  }$ в
дифференциальном
виде%
\begin{align}
dq^{i}  &  =\left\{  q^{i},H_{0}\right\}  dt+\left\{  q^{i},H_{\beta}\right\}
dq^{\beta},\label{dq}\\
dp_{i}  &  =\left\{  p_{i},H_{0}\right\}  dt+\left\{  p_{i},H_{\beta}\right\}
dq^{\beta},\label{dp}\\
\dfrac{\partial H_{\alpha}}{\partial\dot{q}^{\beta}}d\dot{q}^{\beta}+\dfrac
{d}{dt}\left(  \dfrac{\partial H_{0}}{\partial\dot{q}^{\alpha}}+\dfrac
{\partial H_{\beta}}{\partial\dot{q}^{\alpha}}\dot{q}^{\beta}\right)
dt&=\left(  \dfrac{\partial H_{\beta}}{\partial q^{\alpha}}-\dfrac{\partial
H_{\alpha}}{\partial q^{\beta}}+\left\{  H_{\beta},H_{\alpha}\right\}
\right)  dq^{\beta}\nonumber \\&+\left(  \dfrac{\partial H_{0}}{\partial q^{\alpha}}%
-\dfrac{\partial H_{\alpha}}{\partial t}+\left\{  H_{0},H_{\alpha}\right\}
\right)  dt.\label{dhq}%
\end{align}

Осюда видно,
что на $TQ_{2n_{p}}^{\ast}$ мы
получили
уравнения
первого
порядка (\ref{dq})--(\ref{dp})
для
канонических
координат $q^{i}$,
$p_{i}$, как это и
должно быть
(см. (\ref{qp})), в то
время, как на
(неканоническом)
подпространстве
$TQ_{2\left(  n-n_{p}\right)  }$
уравнения (\ref{dhq})
остались
второго
порядка
относительно
неканонических
обобщенных
координат $q^{\alpha}$,
а именно%
\begin{align}
\dot{q}^{i}  &  =\left\{  q^{i},H_{0}\right\}  +\left\{  q^{i},H_{\beta
}\right\}  \dot{q}^{\beta},\label{dq1}\\
\dot{p}_{i}  &  =\left\{  p_{i},H_{0}\right\}  +\left\{  p_{i},H_{\beta
}\right\}  \dot{q}^{\beta},\label{dp1}\\
\dfrac{\partial H_{\alpha}}{\partial\dot{q}^{\beta}}\ddot{q}^{\beta}+\dfrac
{d}{dt}\left(  \dfrac{\partial H_{0}}{\partial\dot{q}^{\alpha}}+\dfrac
{\partial H_{\beta}}{\partial\dot{q}^{\alpha}}\dot{q}^{\beta}\right)   &
=\left(  \dfrac{\partial H_{\beta}}{\partial q^{\alpha}}-\dfrac{\partial
H_{\alpha}}{\partial q^{\beta}}+\left\{  H_{\beta},H_{\alpha}\right\}
\right)  \dot{q}^{\beta}\nonumber \\&+\left(  \dfrac{\partial H_{0}}{\partial q^{\alpha}%
}-\dfrac{\partial H_{\alpha}}{\partial t}+\left\{  H_{0},H_{\alpha}\right\}
\right)  .\label{dhq1}%
\end{align}

Важно
отметить,
что
полученная
система
уравнений
движения (\ref{dq1}%
)--(\ref{dhq1})
частичного
гамильтонового
формализма
справедлива
при любом
числе
редуцированных
импульсов%
\begin{equation}
0\leq n_{p}\leq n,
\end{equation}
то есть не
зависит от
размерности
редуцированного
фазового
пространства.
При этом,
граничные
значения $n_{p}$
соответствуют
лагранжевому
и
гамильтоновому
формализму
соответственно,
так что
имеем три
случая,
которые
описываются
уравнениями
(\ref{dq1})--(\ref{dhq1}):

\begin{enumerate}
\item $n_{p}=0$ ---
лагранжев
формализм на
$TQ_{2n}$ (остается
последнее
уравнение (\ref{dhq1})
без скобок
Пуассона) и
$\alpha=1,\ldots,n$;

\item $0<n_{p}<n$ ---
частичный
гамильтонов
формализм на
$TQ_{2n_{p}}^{\ast}\times TQ_{2\left(  n-n_{p}\right)  }$
(рассматриваются
все уравнения);

\item $n_{p}=n$ ---
стандартный
гамильтонов
формализм на
$TQ_{2n}^{\ast}$ (остаются
первые два
уравнения (\ref{dq1}%
)--(\ref{dp1}) без
вторых
слагаемых,
содержащих
неканонические
обобщенные
скорости),
что
совпадает с
(\ref{qp})) и $i=1,\ldots,n$.
\end{enumerate}

Покажем, что
в случае 1)
получаются
уравнения
Лагранжа
(для
неканонических
переменных
$q^{\alpha}$).
Действительно,
уравнение (\ref{dhq1})
без скобок
Пуассона
(при $n_{p}=0$
канонических
переменных
$q^{i},p_{i}$ вообще
нет)
переписывается
в виде%
\begin{equation}
\dfrac{\partial H_{\alpha}}{\partial\dot{q}^{\beta}}\ddot{q}^{\beta}+\dfrac
{d}{dt}\dfrac{\partial}{\partial\dot{q}^{\alpha}}\left(  H_{0}+H_{\beta}%
\dot{q}^{\beta}\right)  -\dfrac{dH_{\alpha}}{dt}=\dfrac{\partial}{\partial
q^{\alpha}}\left(  H_{0}+H_{\beta}\dot{q}^{\beta}\right)  -\dfrac{\partial
H_{\alpha}}{\partial q^{\beta}}\dot{q}^{\beta}-\dfrac{\partial H_{\alpha}%
}{\partial t},\label{h2}%
\end{equation}
где мы
использовали%
\begin{equation}
\dfrac{d}{dt}\left(  \dfrac{\partial H_{0}}{\partial\dot{q}^{\alpha}}%
+\dfrac{\partial H_{\beta}}{\partial\dot{q}^{\alpha}}\dot{q}^{\beta}\right)
=\dfrac{d}{dt}\left[  \dfrac{\partial}{\partial\dot{q}^{\alpha}}\left(
H_{0}+H_{\beta}\dot{q}^{\beta}\right)  -H_{\beta}\dfrac{\partial}{\partial
\dot{q}^{\alpha}}\dot{q}^{\beta}\right]  =\dfrac{d}{dt}\left[  \dfrac
{\partial}{\partial\dot{q}^{\alpha}}\left(  H_{0}+H_{\beta}\dot{q}^{\beta
}\right)  -H_{\alpha}\right]  .
\end{equation}

Учитывая
выражение
для полной
производной
$dH_{\alpha}\diagup dt$, из (\ref{h2})
получаем%
\begin{equation}
\dfrac{d}{dt}\dfrac{\partial}{\partial\dot{q}^{\alpha}}\left(  H_{0}+H_{\beta
}\dot{q}^{\beta}\right)  =\dfrac{\partial}{\partial q^{\alpha}}\left(
H_{0}+H_{\beta}\dot{q}^{\beta}\right)  .\label{ddh}%
\end{equation}
Формула
определения
предельного
(без
переменных
$q^{i},p_{i}$)
частичного
гамильтониана
(\ref{hl0}) с учетом (\ref{ha})
есть%
\begin{equation}
H_{0}=-H_{\alpha}\dot{q}^{\alpha}-L.
\end{equation}
Отсюда $H_{0}+H_{\beta}\dot{q}^{\beta}%
=-L$, так что из (\ref{ddh})
получаем
уравнения
Лагранжа в
неканоническом
секторе%
\begin{equation}
\dfrac{d}{dt}\dfrac{\partial}{\partial\dot{q}^{\alpha}}L=\dfrac{\partial
}{\partial q^{\alpha}}L.
\end{equation}

Как и в
стандартном
гамильтоновом
формализме
\cite{lan/lif1},
нетривиальная
динамика в
неканоническом
секторе
определяется
наличием
слагаемых со
вторыми
производными,
то есть
присутствием
ненулевых
слагаемых в
левой части
и полных
производных
по времени в
правой части
(\ref{dhq1}).
Рассмотрим
частных
случай
частичного
гамильтонового
формализма,
когда этих
слагаемых
нет, и
назовем его
нединамическим
в
неканоническом
секторе. Для
этого
необходимо
выполнение
условий на
гамильтонианы%
\begin{equation}
\dfrac{\partial H_{0}}{\partial\dot{q}^{\beta}}=0,\ \ \ \ \ \ \dfrac{\partial
H_{\alpha}}{\partial\dot{q}^{\beta}}=0,\ \ \ \ \ \ \alpha,\beta=n_{p}%
+1,\ldots,n.\label{hh}%
\end{equation}
Тогда в (\ref{dhq1})
останется
только
правая
часть,
которую
запишем в
виде%
\begin{equation}
\left(  \dfrac{\partial H_{\beta}}{\partial q^{\alpha}}-\dfrac{\partial
H_{\alpha}}{\partial q^{\beta}}+\left\{  H_{\beta},H_{\alpha}\right\}
\right)  \dot{q}^{\beta}=-\left(  \dfrac{\partial H_{0}}{\partial q^{\alpha}%
}-\dfrac{\partial H_{\alpha}}{\partial t}+\left\{  H_{0},H_{\alpha}\right\}
\right)  ,\label{h1}%
\end{equation}
что
представляет
собой
систему
алгебраических
линейных
уравнений
для
неканонических
скоростей $\dot
{q}^{\alpha}$ при
заданных
гамильтонианах
$H_{0} $, $H_{\alpha}$.
Поскольку в
(\ref{h1})
отсутствуют
неканонические
ускорения $\ddot
{q}^{\alpha}$, то на $TQ_{2\left(  n-n_{p}\right)  }$
при
выполнении
условий (\ref{hh})
нет и
реальной
динамики.
Это
позволяет
провести для
нединамического
в
неканоническом
секторе
частичного
гамильтонового
формализма
следующую аналогию.

\section{Многовременная
динамика}

Условия (\ref{hh})
означают,
что
гамильтонианы
не зависят
явным
образом от
неканонических
скоростей,
то есть $H_{0}=H_{0}\left(  t,q^{i}%
,p_{i},q^{\alpha}\right)  $, $H_{\alpha}=H_{\alpha}\left(  t,q^{i}%
,p_{i},q^{\alpha}\right)  $, $\alpha=n_{p}+1,\ldots,n$.
Таким
образом,
динамическая
задача
определяется
на
многообразии
$TQ_{2n_{p}}^{\ast}\times Q_{\left(  n-n_{p}\right)  }$, так
что $q^{\alpha}$
фактически
играют роль
действительных
параметров,
аналогичных
времени\footnote{В
нединамическом
случае $Q_{\left(  n-n_{p}\right)  }$
изоморфно
реальному
пространству
$R_{\left(  n-n_{p}\right)  }$.}.
Вспоминая
интерпретацию
(\ref{qn1}) и обращая
ее, можно
трактовать
$\left(  n-n_{p}\right)  $
неканонических
обобщенных
координат $q^{\alpha}$
как $\left(  n-n_{p}\right)  $
дополнительных
(к $t$) времен, а $H_{\alpha}
$ как $\left(  n-n_{p}\right)  $
соответствующих
гамильтонианов.
Действительно,
введем
обозначения%
\begin{align}
\tau^{\mu}  &  =t,\ \ \ \ \ \ \mathsf{H}_{\mu}=H_{0},\ \ \ \ \ \ \ \mu=0,\\
\tau^{\mu}  &  =q^{\mu+n_{p}},\ \ \ \ \ \ \mathsf{H}_{\mu}=H_{\mu+n_{p}%
},\ \ \ \ \mu=1,\ldots,\left(  n-n_{p}\right)
\end{align}
где $\mathsf{H}_{\mu}=\mathsf{H}_{\mu}\left(  \tau_{\mu
},q^{i},p_{i}\right)  $ ---
гамильтонианы
многовременной
динамики с $\left(
n-n_{p}+1\right)  $
временами $\tau^{\mu}$.
В\  этой
формулировке
дифференциал
действия
многовременной
динамики $\mathsf{S}%
=\mathsf{S}\left(  \tau^{\mu},q^{i}\right)  $ как
функции
новых времен
запишется в
виде%
\begin{equation}
d\mathsf{S}=p_{i}dq^{i}-\mathsf{H}_{\mu}d\tau^{\mu}.\label{dsp}%
\end{equation}

Отсюда
следует, что
частные
производные
действия $\mathsf{S}$
равны%
\begin{equation}
\dfrac{\partial\mathsf{S}}{\partial q^{i}}=p_{i},\ \ \ \dfrac{\partial
\mathsf{S}}{\partial\tau^{\mu}}=-\mathsf{H}_{\mu},\label{sph}%
\end{equation}
и система $\left(
n-n_{p}+1\right)  $
уравнений
Гамильтона-Якоби
для
многовременной
динамики
имеет вид%
\begin{equation}
\dfrac{\partial\mathsf{S}}{\partial\tau^{\mu}}+\mathsf{H}_{\mu}\left(
\tau_{\mu},q^{i},\dfrac{\partial\mathsf{S}}{\partial q^{i}}\right)
=0,\ \ \ \ \mu=0,\ldots,\left(  n-n_{p}\right)  .\label{s1}%
\end{equation}

Отметим, что
из (\ref{sph}) и (\ref{sl})
следует
дополнительное
соотношение
на $\mathsf{H}_{\mu}$.
Действительно,
продифференцируем
уравнение
Гамильтона-Якоби
(\ref{sl}) по $\tau^{\nu}$%
\begin{align}
\dfrac{\partial^{2}\mathsf{S}}{\partial\tau^{\mu}\partial\tau^{\nu}}  &
=-\dfrac{\partial\mathsf{H}_{\mu}}{\partial\tau^{\nu}}-\dfrac{\partial
\mathsf{H}_{\mu}}{\partial p_{i}}\dfrac{\partial}{\partial\tau^{\nu}}%
\dfrac{\partial\mathsf{S}}{\partial q^{i}}=-\dfrac{\partial\mathsf{H}_{\mu}%
}{\partial\tau^{\nu}}-\dfrac{\partial\mathsf{H}_{\mu}}{\partial p_{i}}\left(
-\dfrac{\partial\mathsf{H}_{\nu}}{\partial q^{i}}-\dfrac{\partial
\mathsf{H}_{\nu}}{\partial p_{j}}\dfrac{\partial}{\partial q^{i}}%
\dfrac{\partial\mathsf{S}}{\partial q^{j}}\right) \nonumber\\
&  =-\dfrac{\partial\mathsf{H}_{\mu}}{\partial\tau^{\nu}}+\dfrac
{\partial\mathsf{H}_{\mu}}{\partial p_{i}}\dfrac{\partial\mathsf{H}_{\nu}%
}{\partial q^{i}}+\dfrac{\partial\mathsf{H}_{\mu}}{\partial p_{i}}%
\dfrac{\partial\mathsf{H}_{\nu}}{\partial p_{j}}\dfrac{\partial^{2}\mathsf{S}%
}{\partial q^{i}\partial q^{j}}.\label{d2s}%
\end{align}
Тогда
антисимметризация
(\ref{d2s}) дает
условия
интегрируемости%
\begin{equation}
\dfrac{\partial^{2}\mathsf{S}}{\partial\tau^{\nu}\partial\tau^{\mu}}%
-\dfrac{\partial^{2}\mathsf{S}}{\partial\tau^{\mu}\partial\tau^{\nu}}%
=\dfrac{\partial\mathsf{H}_{\mu}}{\partial\tau^{\nu}}-\dfrac{\partial
\mathsf{H}_{\nu}}{\partial\tau^{\mu}}+\left\{  \mathsf{H}_{\mu},\mathsf{H}%
_{\nu}\right\}  =0.\label{sth}%
\end{equation}

Чтобы
получить
уравнения
движения,
необходимо
занулить $\delta\mathsf{S}=0 $
вариацию
действия%
\begin{equation}
\mathsf{S}=\int\left(  p_{i}dq^{i}-\mathsf{H}_{\mu}d\tau^{\mu}\right)
\end{equation}
при
независимых
вариациях $\delta q^{i}%
$, $\delta p_{i}$, $\delta\tau^{\mu}$,
исчезающих
на концах
интервала
интегрирования.
Получаем%
\begin{equation}
\delta\mathsf{S}=\int\delta p_{i}\left(  dq^{i}-\dfrac{\partial\mathsf{H}%
_{\mu}}{\partial p_{i}}d\tau^{\mu}\right)  +\int\delta q^{i}\left(
-dp_{i}-\dfrac{\partial\mathsf{H}_{\mu}}{\partial q^{i}}d\tau^{\mu}\right)  ,
\end{equation}
откуда
следуют
уравнения
Гамильтона
для
многовременной
динамики в
дифференциальном
виде \cite{lon/lus/pon}%
\begin{align}
dq^{i}  &  =\left\{  q^{i},\mathsf{H}_{\mu}\right\}  d\tau^{\mu},\label{dq2}\\
dp_{i}  &  =\left\{  p_{i},\mathsf{H}_{\mu}\right\}  d\tau^{\mu},\label{dp2}%
\end{align}
которые
совпадают с
(\ref{dq})--(\ref{dp}). Условия
интегрируемости
(\ref{sth}) можно
также
записать в
дифференциальном
виде\footnote{Как и в
стандартном
случае \cite{arnold},
уравнения (\ref{dq2}%
)--(\ref{hh1}) являются
условиями
замкнутости
дифференциальной
1-формы (\ref{dsp}).}%
\begin{equation}
\left(  \dfrac{\partial\mathsf{H}_{\mu}}{\partial\tau^{\nu}}-\dfrac
{\partial\mathsf{H}_{\nu}}{\partial\tau^{\mu}}+\left\{  \mathsf{H}_{\mu
},\mathsf{H}_{\nu}\right\}  \right)  d\tau^{\nu}=0,\ \ \ \ \mu,\nu
=0,\ldots,\left(  n-n_{p}\right) \label{hh1}%
\end{equation}
что
совпадает с
уравнениями
(\ref{h1}),
записанными
также в
дифференциальном
виде. Таким
образом, мы
показали,
что
нединамический
в
неканоническом
секторе
вариант
частичного
гамильтонового
формализма
(определяемый
уравнениями
движения (\ref{dq1}%
)--(\ref{dhq1}) с
дополнительными
условиями (\ref{hh}))
может быть
сформулирован
как
многовременная
динамика с
числом
времен $\left(  n-n_{p}+1\right)  $
и
уравнениями
(\ref{dq2})--(\ref{hh1}). При этом,
в обеих
формулировках
размерность
фазового
пространства
(и число
обобщенных
импульсов $n_{p}$) произвольна.

\section{Сингулярные
теории}

Рассмотрим
более
подробно
условия (\ref{hh}) и
выразим их в
терминах
лагранжиана.
Используя (\ref{hl0})
и
определение
дополнительных
гамильтонианов
(\ref{ha}), получаем%
\begin{equation}
\dfrac{\partial^{2}L}{\partial\dot{q}^{\alpha}\partial\dot{q}^{\beta}%
}=0,\ \ \ \ \ \ \alpha,\beta=n_{p}+1,\ldots,n.\label{dl}%
\end{equation}
Это
означает,
что динамика
описывается
вырожденным
лагранжианом
(сингулярная
теория), так
что ранг $r_{W}$
матрицы
гессиана%
\begin{equation}
W_{AB}=\left\Vert \dfrac{\partial^{2}L}{\partial\dot{q}^{A}\partial\dot{q}%
^{B}}\right\Vert ,\ \ \ \ \ \ A,B=1,\ldots,n\label{w}%
\end{equation}
не только
меньше
размерности
конфигурационного
пространства
$n$, но и меньше
либо равна
числу
импульсов
(из-за (\ref{dl}))%
\begin{equation}
r_{W}\leq n_{p}.\label{rn}%
\end{equation}

При
рассмотрении
строгого
неравенства
в (\ref{rn}) мы
получаем,
что
определение
\textquotedblleft лишних\textquotedblright%
\ $\left(  n_{p}-r_{W}\right)  $
импульсов
приводит к
появлению $\left(
n_{p}-r_{W}\right)  $ связей, в
точности,
как и в
теории
связей
Дирака \cite{dirac}, где
появляется
$\left(  n-r_{W}\right)  $
(первичных)
связей, если
пользоваться
стандартным
гамильтоновым
формализмом.
Важно, что
размерность
конфигурационного
пространства
$n$ и ранг
матрицы
гессиана $r_{W}$
фиксированы
самой
постановкой
задачи, что
не позволяет
произвольно
изменять
число
связей. В
случае
частичного
гамильтонового
формализма
число
импульсов $n_{p}$
есть
свободный
параметр,
который
может быть
выбран так,
чтобы связи
вообще не
появлялись.
Для этого
естественно
приравнять
число
импульсов
рангу
гессиана%
\begin{equation}
n_{p}=r_{W}.\label{nr1}%
\end{equation}

В
результате,
с одной
стороны,
сингулярная
динамика
(теория с
вырожденным
лагранжианом)
может быть
сформулирована
как
многовременная
динамика с $\left(
n-r_{W}+1\right)  $
временами
(как в
предыдущем
разделе), а с
другой
стороны в
такой
формулировке
не будет
возникать
(первичных, а,
следовательно,
и вторичных,
и более
высокого
уровня) связей
\cite{dup2009,dup2011}.

Для этого,
во-первых,
переименуем
индексы
матрицы
гессиана $W_{AB}$ (\ref{w})
таким
образом,
чтобы
несингулярный
минор ранга
$r_{W}$ находился
в верхнем
левом углу,
при этом
латинскими
буквами $i,j$
обозначим
первые $r_{W}$
индексов, а
греческими
буквами
остальные $\left(
n-r_{W}\right)  $ индексов
$\alpha,\beta$. Далее,
запишем
уравнения
движения (\ref{dq1}%
)-(\ref{dp1}), (\ref{h1}) в виде%
\begin{align}
&  \dot{q}^{i}=\left\{  q^{i},H_{0}\right\}  +\left\{  q^{i},H_{\beta
}\right\}  \dot{q}^{\beta},\label{dq3}\\
&  \dot{p}_{i}=\left\{  p_{i},H_{0}\right\}  +\left\{  p_{i},H_{\beta
}\right\}  \dot{q}^{\beta},\label{dp3}\\
&  F_{\alpha\beta}\dot{q}^{\beta}=G_{\alpha},\label{fq}%
\end{align}
где значения
индексов
связаны с
рангом
гессиана $i=1,\ldots,r_{W}$,
$\alpha,\beta=r_{W}+1,\ldots,n$, и%

\begin{align}
F_{\alpha\beta}  &  =\dfrac{\partial H_{\alpha}}{\partial q^{\beta}}%
-\dfrac{\partial H_{\beta}}{\partial q^{\alpha}}+\left\{  H_{\alpha},H_{\beta
}\right\}  ,\label{f}\\
G_{\alpha}  &  =D_{\alpha}H_{0}=\dfrac{\partial H_{0}}{\partial q^{\alpha}%
}-\dfrac{\partial H_{\alpha}}{\partial t}+\left\{  H_{0},H_{\alpha}\right\}
.\label{dh}%
\end{align}

Отметим, что
система
уравнений (\ref{dq3}%
)--(\ref{dh}) совпадает
с
уравнениями,
полученными
в подходе к
сингулярным
теориям,
использующем
смешанные
решения
уравнения
Клеро \cite{dup2009,dup2011} (за
исключением
слагаемого с
производной
$H_{\alpha}$ по времени
в (\ref{dh})).

Уравнения (\ref{dq3}%
)--(\ref{fq})
представляют
собой
систему
дифференциальных
уравнений
первого
порядка для
канонических
координат $q^{i},p_{i}$,
в то время,
как
относительно
неканонических
скоростей $\dot
{q}^{\alpha}$ --- это
алгебраическая
система.
Действительно,
(\ref{fq}) есть
обычная
система
линейных
уравнений
относительно
$\dot{q}^{\alpha}$, по
свойствам
решений
которой
можно
классифицировать
классические
сингулярные
теории.
Будем
рассматривать
только те
случаи,
когда
система (\ref{fq})
совместна,
тогда
имеется две
возможности,
определяемые
рангом
антисимметричной
матрицы $F_{\alpha\beta}$

\begin{enumerate}
\item
Некалибровочная
теория,
когда $\operatorname*{rank}F_{\alpha\beta}%
=r_{F}=n-r_{W}$ полный,
так что
матрица $F_{\alpha\beta}$
обратима.
Тогда из (\ref{fq})
можно
определить
все
неканонические
скорости
\begin{equation}
\dot{q}^{\alpha}=\bar{F}^{\alpha\beta}G_{\beta},\label{q}%
\end{equation}
где $\bar{F}^{\alpha\beta}$ ---
матрица,
обратная к
$F_{\alpha\beta}$,
определяемая
уравнением
$\bar{F}^{\alpha\beta}F_{\beta\gamma}=F_{\gamma\beta}\bar{F}^{\beta\alpha
}=\delta_{\gamma}^{\alpha}$.

\item
Калибровочная
теория,
когда ранг
$F_{\alpha\beta}$ неполный,
то есть $r_{F}<n-r_{W}$, и
матрица $F_{\alpha\beta}$
необратима.
В этом
случае из (\ref{fq})
находятся
только $r_{F}$
неканонических
скоростей, в
то время, как
$\left(  n-r_{W}-r_{F}\right)  $
скоростей
остаются
произвольными
калибровочными
параметрами,
которые
соответствуют
симметриям
сингулярной
динамической
системы. В
частном
случае $r_{F}=0$ (или
нулевой
матрицы $F_{\alpha\beta}$)
из (\ref{fq})
получаем%
\begin{equation}
G_{\alpha}=0,
\end{equation}
и все
неканонические
скорости
являются $\left(
n-r_{W}\right)  $
калибровочными
параметрами теории.
\end{enumerate}

В первом
случае
(некалибровочной
теории)
можно
исключить
все
неканонические
скорости с
помощью (\ref{q}) и
подставить в
(\ref{dq3})--(\ref{dp3}). Тогда
получаем
уравнения
типа
Гамильтона
для
некалибровочной
сингулярной
системы%
\begin{align}
\dot{q}^{i}  &  =\left\{  q^{i},H_{0}\right\}  _{nongauge},\label{qnon}\\
\dot{p}_{i}  &  =\left\{  p_{i},H_{0}\right\}  _{nongauge},\label{pnon}%
\end{align}
где мы ввели
новую
(некалибровочную)
скобку для
двух
динамических
величин $A,B$%
\begin{equation}
\left\{  A,B\right\}  _{nongauge}=\left\{  A,B\right\}  +D_{\alpha}A\cdot
\bar{F}^{\alpha\beta}\cdot D_{\beta}B,\label{nong}%
\end{equation}
и $D_{\alpha}$
определено в
(\ref{dh}). Из (\ref{qnon})--(\ref{pnon})
следует, что
новая
некалибровочная
скобка (\ref{nong})
однозначно
определяет
эволюцию
любой
динамической
величины $A$ от
времени%
\begin{equation}
\dfrac{dA}{dt}=\dfrac{\partial A}{\partial t}+\left\{  A,H_{0}\right\}
_{nongauge}.\label{da}%
\end{equation}
Важно, что
некалибровочная
скобка (\ref{nong})
обладает
всеми
свойствами
скобки
Пуассона:
она
антисимметрична
и
удовлетворяет
тождеству
Якоби.
Поэтому
определение
(\ref{nong}) может
рассматриваться
как
некоторая
деформация
скобки
Пуассона, но
только не
для всех $2n$
переменных,
как в
стандартном
случае, а
только для $2r_{W}$
канонических
$\left(  q^{i},p_{i}\right)  $, $i=1,\ldots,r_{W}$. Из
(\ref{nong}) и (\ref{da})
следует, что,
как и в
стандартном
случае, если
$H_{0}$ не зависит
явно от
времени, то
он сохраняется.

Во втором
случае
(калибровочной
теории)
можно
исключить
только часть
неканонических
скоростей $\dot
{q}^{\alpha}$, число
которых
равно рангу
$r_{F}$ матрицы $F_{\alpha\beta}$,
а остальные
скорости
остаются
произвольными
и могут
служить
калибровочными
параметрами.
Действительно,
если матрица
$F_{\alpha\beta}$
сингулярна и
имеет ранг $r_{F}$,
то можно
снова
привести ее
к такому
виду, что
несингулярный
минор
размером $r_{F}\times r_{F}$
будет
находиться в
левом
верхнем
углу. Тогда в
системе (\ref{fq})
только
первые $r_{F}$
уравнений
будут
независимы.
Представим
(\textquotedblleft%
неканонические\textquotedblright%
) индексы $\alpha,\beta
=r_{W}+1,\ldots,n$ в виде пар
$\left(  \alpha_{1},\alpha_{2}\right)  $, $\left(  \beta_{1},\beta_{2}\right)
$, где $\alpha_{1},\beta_{1}=r_{W}+1,\ldots,r_{F}$
нумеруют
первые $r_{F}$
независимых
строк
матрицы $F_{\alpha\beta}$ и
соответствуют
ее
несингулярному
минору $F_{\alpha_{1}\beta_{1}}$,
остальные $\left(
n-r_{W}-r_{F}\right)  $ строк
будут
зависимы от
первых, и $\alpha_{2},\beta
_{2}=r_{F}+1,\ldots,n$. Тогда
система (\ref{fq})
может быть
записана в
виде%
\begin{align}
F_{\alpha_{1}\beta_{1}}\dot{q}^{\beta_{1}}+F_{\alpha_{1}\beta_{2}}\dot
{q}^{\beta_{2}}  &  =G_{\alpha_{1}},\label{fg1}\\
F_{\alpha_{2}\beta_{1}}\dot{q}^{\beta_{1}}+F_{\alpha_{2}\beta_{2}}\dot
{q}^{\beta_{2}}  &  =G_{\alpha_{2}}.\label{fg2}%
\end{align}
Поскольку $F_{\alpha
_{1}\beta_{1}}$
несингулярна
по
построению,
мы можем
выразить
первые $r_{F}$
неканонических
скоростей $\dot
{q}^{\alpha_{1}}$ через
остальные $\left(
n-r_{W}-r_{F}\right)  $
скорости $\dot{q}^{\alpha_{2}%
}$%
\begin{equation}
\dot{q}^{\alpha_{1}}=\bar{F}^{\alpha_{1}\beta_{1}}G_{\beta_{1}}-\bar
{F}^{\alpha_{1}\beta_{1}}F_{\beta_{1}\alpha_{2}}\dot{q}^{\alpha_{2}%
},\label{q1}%
\end{equation}
где $\bar{F}^{\alpha_{1}\beta_{1}}$ --- $r_{F}\times r_{F}%
$-матрица,
обратная к
$F_{\alpha_{1}\beta_{1}}$. Далее,
из-за того,
что $\operatorname*{rank}F_{\alpha_{1}\beta_{1}}=r_{F}$,
можно
выразить
остальные
блоки через
несингулярный
блок $F_{\alpha_{1}\beta_{1}}$
следующим
образом%
\begin{align}
F_{\alpha_{2}\beta_{1}}  &  =\lambda_{\alpha_{2}}^{\alpha_{1}}F_{\alpha
_{1}\beta_{1}},\label{f1}\\
F_{\alpha_{2}\beta_{2}}  &  =\lambda_{\alpha_{2}}^{\alpha_{1}}F_{\alpha
_{1}\beta_{2}}=\lambda_{\alpha_{2}}^{\alpha_{1}}\lambda_{\beta_{2}}%
^{\gamma_{1}}F_{\alpha_{1}\gamma_{1}},\label{f2}\\
G_{\alpha_{2}}  &  =\lambda_{\alpha_{2}}^{\alpha_{1}}G_{\alpha_{1}},\label{g}%
\end{align}
где $\lambda_{\alpha_{2}}^{\alpha_{1}}=\lambda_{\alpha_{2}%
}^{\alpha_{1}}\left(  q^{i},p_{i},q^{\alpha}\right)  $ --- $r_{F}\times\left(
n-r_{W}-r_{F}\right)  $ гладких
функций.
Поскольку
матрица $F_{\alpha\beta}$
задана, то мы
можем
определить
все функции
$\lambda_{\alpha_{2}}^{\alpha_{1}}$ из $r_{F}\times\left(
n-r_{W}-r_{F}\right)  $
уравнений (\ref{f1})%
\begin{equation}
\lambda_{\alpha_{2}}^{\alpha_{1}}=F_{\alpha_{2}\beta_{1}}\bar{F}^{\alpha
_{1}\beta_{1}}.
\end{equation}

Из-за того,
что $\left(  n-r_{W}-r_{F}\right)  $
скоростей $\dot
{q}^{\alpha_{2}}$
произвольны,
мы можем
положить их
равными нулю%
\begin{equation}
\dot{q}^{\alpha_{2}}=0,\ \ \ \ \ \alpha_{2}=r_{F}+1,\ldots,n,\label{q2}%
\end{equation}
что можно
считать
некоторым
калибровочным
условием.
Тогда из (\ref{q1})
следует, что%
\begin{equation}
\dot{q}^{\alpha_{1}}=\bar{F}^{\alpha_{1}\beta_{1}}G_{\beta_{1}},\ \ \ \ \alpha
_{1}=r_{W}+1,\ldots,r_{F}.
\end{equation}

По аналогии
с (\ref{nong}) введем
новые
(калибровочные)
скобки%
\begin{equation}
\left\{  A,B\right\}  _{gauge}=\left\{  A,B\right\}  +D_{\alpha_{1}}A\cdot
\bar{F}^{\alpha_{1}\beta_{1}}\cdot D_{\beta_{1}}B.\label{gaug}%
\end{equation}

Тогда
уравнения
движения (\ref{dq3}%
)--(\ref{fq}) запишутся
в
гамильтоновом
виде, как и (\ref{qnon}%
)--(\ref{pnon})%
\begin{align}
\dot{q}^{i}  &  =\left\{  q^{i},H_{0}\right\}  _{gauge},\\
\dot{p}_{i}  &  =\left\{  p_{i},H_{0}\right\}  _{gauge}.
\end{align}

Эволюция
физической
величины $A$ во
времени, как
и (\ref{da}), также
определяется
калибровочной
скобкой (\ref{gaug})%
\begin{equation}
\dfrac{dA}{dt}=\dfrac{\partial A}{\partial t}+\left\{  A,H_{0}\right\}
_{gauge}.
\end{equation}

В частном
предельном
случае
нулевого
ранга $r_{F}=0$
имеем%
\begin{equation}
F_{\alpha\beta}=0,\label{f0}%
\end{equation}
а,
следовательно,
все
дополнительные
гамильтонианы
зануляются
$H_{\alpha}=0$, тогда из
определения
(\ref{ha}) видно, что
лагранжиан
не зависит
от
неканонических
скоростей $\dot
{q}^{\alpha}$ и поэтому
с учетом (\ref{f0}) из
(\ref{fq}) получаем,
что и
частичный
гамильтониан
$H_{0}$ не зависит
от
неканонических
обобщенных
координат $q^{\alpha}$%
\begin{equation}
\dfrac{\partial H_{0}}{\partial q^{\alpha}}=0
\end{equation}
при условии
независимости
$H_{0}$ от времени
явно. В этом
случае
калибровочные
скобки
совпадают со
скобками
Пуассона,
поскольку
второе
слагаемое в
(\ref{gaug}) зануляется.

Таким
образом, мы
показали,
что
сингулярные
теории (с
вырожденным
лагранжианом)
на
классическом
уровне могут
быть описаны
в рамках
частичного
гамильтонового
формализма с
числом
импульсов $n_{p}$,
равном рангу
$r_{W}$ матрицы
гессиана $n_{p}=r_{W}$
или
сформулированы
как
многовременная
динамика с
числом
времен $\left(  n-r_{W}+1\right)  $
без введения
дополнительных
соотношений
на
динамические
переменные (связей).

\section{Причины
возникновения
связей}

Как было
отмечено
выше (после (\ref{rn})),
введение
дополнительных
динамических
переменных с
необходимостью
должно
приводить к
появлению
дополнительных
соотношений
на них.
Например,
введем в
рассмотрение
\textquotedblleft лишние\textquotedblright%
\  импульсы $p_{\alpha}$
(поскольку
мы получили
полное
описание
динамики и
без них),
которые
соответствуют
неканоническим
обобщенным
скоростям $\dot
{q}^{\alpha}$ по
стандартному
определению
\cite{dirac}%
\begin{equation}
p_{\alpha}=\dfrac{\partial L}{\partial\dot{q}^{\alpha}},\ \ \ \ \ \alpha
=r_{W}+1,\ldots,n,\label{pa}%
\end{equation}
так что (\ref{pa})
вместе с
определением
канонических
обобщенных
импульсов (\ref{pp})
совпадают с
определением
\textquotedblleft полных\textquotedblright%
\  импульсов (\ref{p}).
Пользуясь
определением
дополнительных
гамильтонианов
(\ref{ha}), получаем
столько же $\left(
n-r_{W}\right)  $
соотношений%
\begin{equation}
\Phi_{\alpha}=p_{\alpha}+H_{\alpha}=0,\ \ \ \ \ \alpha=r_{W}+1,\ldots
,n,\label{fa}%
\end{equation}
которые
называются
(первичными)
связями \cite{dirac} (в
разрешенном
виде). Эти
соотношения
напоминают
процедуру
расширения
фазового
пространства
(\ref{pn1}). Можно
ввести любое
количество
$n_{p}^{\left(  add\right)  }$ \textquotedblleft%
лишних\textquotedblright%
\  импульсов $0\leq
n_{p}^{\left(  add\right)  }\leq n-r_{W}$, тогда
в теории
появится
столько же
$n_{p}^{\left(  add\right)  }$
(первичных)
связей. В
частичном
гамильтоновом
формализме
нами был
рассмотрен
случай $n_{p}^{\left(  add\right)  }=0$,
в то время
как в теории
Дирака $n_{p}^{\left(  add\right)
}=n-r_{W}$, хотя можно
взять и
промежуточные
варианты,
что
обусловливается
конкретной задачей.

Переход к
гамильтониану
по
стандартный
формуле%
\begin{equation}
H_{total}=p_{i}\dot{q}^{i}+p_{\alpha}\dot{q}^{\alpha}-L,\label{ht}%
\end{equation}
напрямую
невозможен,
поскольку
нельзя
выразить
неканонические
скорости $\dot{q}^{\alpha}$
через \textquotedblleft%
лишние\textquotedblright%
\  импульсы $p_{\alpha}$
и далее
применить
преобразование
Лежандра. Но
можно
преобразовать
$H_{total}$ (\ref{ht}) таким
образом,
чтобы
воспользоваться
методом
неопределенных
коэффициентов.
Здесь важно,
что связи $\Phi_{\alpha}$
не зависят
от
обобщенных
скоростей $\dot
{q}^{\alpha}$, как и
гамильтонианы
$H_{0},H_{\alpha}$, из-за
того, что
ранг матрицы
гессиана
равен $r_{W}$.
Поэтому
(полный)
гамильтониан
можно
записать%
\begin{equation}
H_{total}=H_{0}+\dot{q}^{\alpha}\Phi_{\alpha},\label{ht1}%
\end{equation}
где $\dot{q}^{\alpha}$
играют роль
неопределенных
коэффициентов.
В терминах
полного
гамильтониана
и полной
скобки
Пуассона (\ref{abf})
уравнения
движения
запишутся в
гамильтоновом
виде%
\begin{align}
dq^{A}  & =\left\{  q^{A},H_{total}\right\}  _{full}\ dt,\label{qpf}\\
dp_{A}  & =\left\{  p_{A},H_{total}\right\}  _{full}\ dt\label{qpf1}%
\end{align}
с учетом $\left(  n-r_{W}\right)  $
дополнительных
условий (\ref{fa}).
Однако
уравнений (\ref{qpf}%
)--(\ref{qpf1}) и (\ref{fa})
недостаточно
для решения
задачи:
необходимы
еще
уравнения
для
нахождения
неопределенных
коэффициентов
$\dot{q}^{\alpha}$ в (\ref{ht1}). Такие
уравнения
можно
получить из
некоторого
дополнительного
принципа,
например,
сохранения
связей (\ref{fa}) во
времени \cite{dirac}%
\begin{equation}
\dfrac{d\Phi_{\alpha}}{dt}=0.
\end{equation}
Зависимость
от времени
любой
физической
величины $A$
теперь
определяется
полным
гамильтонианом
и полной
скобкой
Пуассона%
\begin{equation}
\dfrac{dA}{dt}=\dfrac{\partial A}{\partial t}+\left\{  A,H_{total}\right\}
_{full}.\label{da1}%
\end{equation}
Если связи
не зависят
явно от
времени, то
из (\ref{da1}) и (\ref{ht1})
получаем%
\begin{equation}
\left\{  \Phi_{\alpha},H_{total}\right\}  _{full}=\left\{  \Phi_{\alpha}%
,H_{0}\right\}  _{full}+\left\{  \Phi_{\alpha},\Phi_{\beta}\right\}
_{full}\ \dot{q}^{\beta}=0,\label{fh}%
\end{equation}
что
представляет
собой
систему
уравнений
для
нахождения
неопределенных
коэффициентов
$\dot{q}^{\alpha}$ и
совпадает с
(\ref{fq}), поскольку%
\begin{align}
F_{\alpha\beta}  & =\left\{  \Phi_{\alpha},\Phi_{\beta}\right\}
_{full},\label{fdh}\\
D_{\alpha}H_{0}  & =\left\{  \Phi_{\alpha},H_{0}\right\}  _{full}.\label{fdh1}%
\end{align}
Однако в
отличе от
сокращенного
описания
(без $\left(  n-r_{W}\right)  $ \textquotedblleft%
лишних\textquotedblright%
\  импульсов $p_{\alpha}%
$), когда (\ref{fq})
предствляет
собой
систему $\left(  n-r_{W}\right)  $
линейных
уравнений
относительно
$\left(  n-r_{W}\right)  $
неизвестных
$\dot{q}^{\alpha}$,
расширенная
система (\ref{fh})
может
приводить
еще и к
дополнительным
связям
(высших
этапов), что
существенно
усложняет
анализ
физической
динамики \cite{sundermeyer}.
Из (\ref{fdh})--(\ref{fdh1})
следует, что
новые скобки
(калибровочные
(\ref{gaug}) и
некалибровочные
(\ref{nong})) переходят
в
соответствующие
скобки
Дирака.
Отметим
также, что
наша
классификация
на
калибровочные
и
некалибровочные
теории
соответствует
связям
первого и
второго
класса (рода)
\cite{dirac}, а
предельный
случай $F_{\alpha\beta}=0$ (\ref{f0})
отвечает
абелевым
связям \cite{gog/khv/per,lor05}.

\section{Выводы}

Таким
образом, в
работе
построена
\textquotedblleft%
сокращенная\textquotedblright%
\  формулировка
сингулярных
классических
теорий, в
рамках
которой не
возникает
понятия
связей,
поскольку не
вводится \textquotedblleft%
лишних\textquotedblright%
\  динамических
переменных,
а именно,
обобщенных
импульсов,
соответствующих
неканоническим
координатам.
В этих целях
строится
частичный
гамильтонов
формализм и
показывается,
что его
частный
случай
эффективно
описывает
многовременную
динамику.
Доказано,
что
сингулярные
теории (с
вырожденным
лагранжианом)
описываются
в рамках
этих двух
подходов без
введения
дополнительных
соотношений
между
динамическими
величинами
(связей), если
число
канонических
обобщенных
импульсов
совпадает с
рангом
матрицы
гессиана $n_{p}=r_{W}$,
то есть в
редуцированном
фазовом
пространстве.
С физической
точки
зрения, в
самом
введении \textquotedblleft%
лишних\textquotedblright%
\  импульсов
нет
необходимости,
поскольку в
этих
(вырожденных)
направлениях
нет динамики.

Гамильтонова
формулировка
сингулярных
теорий
проведена с
помощью
новых скобок
(калибровочных
(\ref{gaug}) и
некалибровочных
(\ref{nong})), которые
обладают
всеми
свойствами
скобок
Пуассона
(антисимметричность,
удовлетворение
тождеству
Якоби и
запись через
них
уравнений
движения и
эволюции
системы во
времени). При
расширении
фазового
пространства
до полного
эти скобки
переходят в
скобки
Дирака, а на
\textquotedblleft лишние\textquotedblright%
\  импульсы
накладываются
связи.

Проведенный
анализ
позволяет
предположить,
что
квантование
сингулярных
систем в
рамках
предлагаемого
\textquotedblleft%
сокращенного\textquotedblright%
\  подхода
может быть
проведено
стандартным
способом \cite{dirac},
но
квантоваться
будут не все $2n$
переменных
расширенного
фазового
пространства,
а только $2r_{W}$
переменных
редуцированного
фазового
пространства.
Остальные
(неканонические)
переменные
рассматриваются
в качестве
непрерывных параметров.

\bigskip

\noindent Автор
выражает
благодарность
А.А. Воронову,
У. Гюнтеру,
Г.Ч.
Куринному,
А.Ю.
Нурмагамбетову,
Дж. Сташефу
за полезные обсуждения.

\bigskip

\begin{small}

\end{small}

\end{document}